%% file: arxiv-v3.tex
\documentclass[11pt]{article}

\input{pre}

\begin{document}
	
  \begin{center}
    {\Large \bf \sffamily On the consistency of non-commutative geometry inspired Reissner-Nordstr\"{o}m black hole solution}
  \end{center}

  \begin{center}
    \vspace{10pt}

    {{\bf \sffamily G{\"o}khan Alka\c{c},}${}^{a}\,${\bf \sffamily  Murat Mesta}${}^{b}\,$ {\bf \sffamily and G\"{o}n\"{u}l \"{U}nal}${}^{c}$}
    \\[4mm]

    {\small 
    {\it ${}^a$Department of Aerospace Engineering, Faculty of Engineering,\\ At{\i}l{\i}m University, 06836 Ankara, T\"{u}rkiye}\\[2mm]

    {\it ${}^b$Department of Electrical and Electronics Engineering, Faculty of Engineering,\\ At{\i}l{\i}m University, 06836 Ankara, T\"{u}rkiye}\\[2mm]

    {\it ${}^c$Department of Biomedical Engineering, Faculty of Engineering,\\ Başkent University, 06790 Ankara, T\"{u}rkiye}\\[2mm]

    {\it E-mail:} {\mail{alkac@mail.com}, \mail{murat.mesta@atilim.edu.tr}, \mail{gunalmesta@baskent.edu.tr}}
    }
    \vspace{2mm}
  \end{center}

  \centerline{{\bf \sffamily Abstract}}
  \vspace*{1mm}
\noindent We revisit the non-commutative geometry inspired Reissner-Nordstr\"{o}m
black hole solution obtained by smearing the point sources with a Gaussian
distribution. We show that while the form of the metric function and the physical
properties derived from that remain valid, not all the components of Einstein
equations are satisfied: the angular components fail by exactly the Lorentz force
density exerted by the electric field on the smeared charge, which the standard
electromagnetic energy-momentum tensor leaves unbalanced. We construct an improved
energy-momentum tensor that consistently satisfies Einstein equations. It is the
unique conserved completion: the standard Maxwell tensor supplemented by an
interaction stress supported on the smeared source. The improved tensor leads to a
different prediction for the region where the strong energy condition is violated.
For certain choice of parameters, our proposal predicts the violation of the energy
condition outside the Cauchy horizon, which might be relevant for the stability of
the inner horizon and strong cosmic censorship.
  \par\noindent\rule{\textwidth}{0.5pt}

\vspace{3mm}
Non-commutativity is quite an old idea and was first considered by Snyder \cite{Snyder:1946qz, Snyder:1947nq} as a possible cure to the divergences in quantum field theories. Interest in the topic naturally declined after the development of systematic renormalization techniques. In the 1980's, non-commutative geometry was studied extensively by mathematicians as a generalization of differential geometry \cite{Connes1985,Woronowicz1987a,Woronowicz1987b,Connes1994}. However, it did not attract much attention in the high energy physics community until 1990s, when important results from the study of open strings appeared \cite{Witten:1995im, Seiberg:1999vs}. In the presence of a constant Neveu-Schwarz $B$-field, target spacetime coordinates become non-commutative. In the low energy limit, this implies the existence of non-commutative field theories, i.e., field theories defined on non-commutative space-time.

The non-commutativity of spacetime coordinates
\begin{equation}\label{com}
	\left[x^\mu, x^\nu\right]=i \theta^{\mu \nu},
\end{equation}
where $\theta^{\mu \nu}$ is an anti-symmetric tensor, directly makes the notion of a point meaningless due to the uncertainty relation 
\begin{equation}
	\Delta x^\mu \Delta x^\nu \geq \frac{1}{2}\left|\theta^{\mu \nu}\right|.
\end{equation}
This suggests that the non-commutativity can be used in a classical theory for the regularization of singularities. The commutator in \eqref{com} can be imposed on fields in a Lagrangian by replacing the point-wise multiplication by the Groenewold-Moyal $\star$ product \cite{Groenewold:1946kp, Moyal:1949sk}
\begin{equation}\label{moyal}
	(f \star g)(x)=\left.e^{\frac{i}{2} \theta^{\mu \nu} \partial_\mu \partial_\nu} f(x) g(y)\right|_{y \rightarrow x}.
\end{equation}
Involving infinitely many derivatives, the $\star$ product is inherently non-local. As a result, finding an exact solution to field equations is quite difficult. While one might try to study the effect of non-commutativity perturbatively, truncating the series in \eqref{moyal} leads to a local theory, and therefore, the non-locality is lost. In order to obtain non-commutative deformations of the solutions of general relativity, one might try the same strategy. However, solutions obtained from a truncated version of the non-commutative Einstein-Hilbert action generated by the $\star$ product cannot realize the desired non-local behavior, and therefore removing singularities is unlikely.

Fortunately, an alternative route has been established thanks to the results of Smailagic and Spallucci \cite{Smailagic:2003yb, Smailagic:2003rp} obtained by employing coherent states. This approach is necessary since there are no common eigenstates of the coordinates when they do not commute. In their formulation, the non-commutativity introduces an exponential cut-off into the propagator of a free particle at large momenta. As a result, the Dirac delta distribution describing point sources are modified as follows
\begin{equation}\label{subs}
\d(\vec{r}) \to \frac{e^{-r^2 /4 \theta}}{(4 \pi \theta)^{(D-1) / 2}}, 
\end{equation}
where $D$ is the dimension of the spacetime. The point source is smeared by a Gaussian distribution with a width $\sqrt{\th}$, where $\th$ is a constant with dimension length squared. One now has a single parameter $\th$ characterizing the non-commutative effects because in order to preserve Lorentz invariance and unitarity, one should take the anti-symmetric tensor in \eqref{com} as $\theta^{\mu \nu}=\theta \operatorname{diag}\left(\epsilon_{i j}, \epsilon_{i j}, \ldots\right)$ \cite{Smailagic:2004yy}.

In gravitational theories, one may adopt an effective approach and use the substitution rule in \eqref{subs} to probe the effects of non-commutativity while keeping the theories in their standard, commutative form. In this way, a large number of non-commutative geometry inspired black hole solutions of general relativity has been obtained including the Schwarzschild black hole \cite{Nicolini:2005vd}, the Reissner-Nordstr\"{o}m black hole \cite{Ansoldi:2006vg}, their higher-dimensional generalizations \cite{Rizzo:2006zb, Spallucci:2008ez} and the Banados-Teitelboim-Zanelli black hole in three spacetime dimensions \cite{Rahaman:2013gw} (see \cite{Nicolini:2008aj} for a review). It is also possible to apply this idea beyond general relativity. For example, the non-commutative geometry inspired black hole solution of 5d Einstein-Gauss-Bonnet theory was given in \cite{Ghosh:2017odw}. Arguably, the most important results of the non-commutative effects are the modifications obtained in black hole thermodynamics \cite{Myung:2006mz, Banerjee:2008du, Kim:2008vi, Nozari:2008rc}. For the Schwarzschild black hole, the end point of the black hole evaporation is an extremal black hole with a zero temperature and one ends up with a regular de Sitter core \cite{Nicolini:2005vd}. This is a striking difference compared to the standard scenario, where the temperature and the curvature of the final state diverges. As expected, the non-commutativity solves the singularity problems.

In order to study the non-commutative effects on the Reissner-Nordstr\"{o}m black hole, the substitution in \eqref{subs} should also be used for the point electric charge \cite{Ansoldi:2006vg, Spallucci:2008ez}. After obtaining the modified form of the electric field configuration, the field equations are solved with the modified energy momentum tensor. In this letter, we notice that only the $tt$-component of the field equations is solved through this procedure, and therefore an extra modification of the energy-momentum tensor is needed to solve the other components consistently. We identify the failure of the remaining components in closed form as the unbalanced Lorentz force density exerted by the electric field on the smeared charge, and we show that the required modification is the unique conserved completion of the standard electromagnetic energy-momentum tensor. We would like to emphasize that we do \emph{not} contradict the previously obtained charged solutions or the implications on black hole thermodynamics, apart from the energy-momentum tensor. However, our improvement of the energy-momentum tensor is important in some physical applications, for example, a different value is obtained for the critical radius $r_*$ that determines the region ($r<r_*$) where the strong energy condition is violated.

To set the stage, let us review the non-commutative geometry inspired Schwarzschild black hole \cite{Nicolini:2005vd}. Taking a line element of the following form
\begin{equation}\label{met}
    ds^2 = -f(r)dt^2+\frac{dr^2}{f(r)}+r^2d\O^2,
\end{equation}
components of the Einstein tensor can be calculated as
\begin{align}
	 G^\mu_{\,\nu} &=\text{diag} \left\{\psi,\psi,\psi+\frac{r}{2}\psi^\pr,\psi+\frac{r}{2}\psi^\pr\right\},\label{Einsteintensor}\\
	\psi &= \frac{1}{r^2}(-1+f+rf^{\pr}),
\end{align}
where prime denotes the derivative with respect to the radial coordinate $r$.
Note that all the other components can be written in terms of the $tt$-component.

For the consistency of the Einstein equations
\begin{equation}
    G^\mu_{\ \nu} = 8\pi T^{\mu}_{\ \nu},
\end{equation}
the energy-momentum tensor should be of the following form
\begin{align}
    T^{\mu}_{\ \nu}&=\text{diag}\left\{-\e,p_r,p_\perp,p_\perp\right\},\label{Tmatt}\\
    p_r&=-\e, \qquad p_\perp=-\e-\frac{r}{2}\e^\pr,\label{ptoeps}
\end{align}
where $\e$ is the energy density, $p_r$ and $p_\perp$ are the radial and transverse pressure components respectively\footnote{The relation between the pressure components and the energy density in \eqref{ptoeps} can also be derived from the covariant conservation of the energy momentum tensor, i.e. $\nabla_\mu T^{\mu}_{\ \nu}=0$. Here, we derive it by observing the relation between the components of the Einstein tensor for spacetimes with the metric in \eqref{met}.}.

Taking a point mass at the origin as $\e=M\d(\vec{r})$ and defining the pressure components in accordance with \eqref{ptoeps}, one obtains a consistent set of field equations. In this case, it is enough to solve only the $tt$-component and one obtains the metric function of the ordinary Schwarzschild black hole 
\begin{equation}
    f(r)=1-\frac{2M}{r}.
\end{equation}
For the noncommutative geometry inspired version, we need to modify the energy-momentum tensor by using the substitution rule \eqref{subs}. The new energy-momentum tensor is given by
\begin{align}
    &T^\mu_{\ \nu}|_\text{matt.}=\text{diag}\left\{-\e|_\text{matt.},p_r|_\text{matt.},p_\perp|_\text{matt.},p_\perp|_\text{matt.}\right\},\label{Tmatt2}\\
    \e|_\text{matt.}=\frac{M}{(4\pi\th)^{3/2}}&e^{-r^2/4\th}, \quad p_r|_\text{matt.}=-\e|_\text{matt.}, \quad p_\perp|_\text{matt.}=-\e|_\text{matt.}-\frac{r}{2}\e|_\text{matt.}^\pr.\label{Tmatt3}
\end{align}
The solution can again be found from the $tt$-component of Einstein equations
\begin{equation}
    G^\mu_{\ \nu} = 8\pi T^\mu_{\ \nu}|_\text{matt.},
\end{equation}
which reads
\begin{equation}
    f(r)=1-\frac{4M}{\sqrt{\pi}r}\g(3/2,r^2/4\th),
\end{equation}
where $\g(s,x)=\int\limits_0^x t^{s-1}e^{-t}\,\dd{t} $ is the lower incomplete gamma function.

We can now turn our attention to Einstein-Maxwell theory whose field equations are given by
\begin{align}
	&G^\mu_{\ \nu}=8\pi\left(T^\mu_{\ \nu}|_\text{matt.}+T^\mu_{\ \nu}|_\text{el.}\right),\label{EM}\\
	&\pd_\mu\left(\sqrt{-g}F^{\m\n}\right)=\sqrt{-g}J^\n.\label{max}
\end{align}
Note that there are now two contributions to the energy-momentum tensor. $T^\mu_{\ \nu}|_\text{matt.}$ is the contribution from the point mass that we discussed in the Schwarzschild case. $T^\mu_{\ \nu}|_\text{el.}$ represents the contribution due to the electromagnetic fields and it is given by
\begin{equation}\label{Telgen}
    T^\mu_{\ \nu}|_\text{el.}=F^{\m\a}F_{\n\a}-\frac{1}{4}\d^\m_{\ \n}F_{\a\b}F^{\a\b}.
\end{equation}
Assuming the following form for the vector potential
\begin{equation}
    A=\phi(r)dt,\label{gauge}
\end{equation}
and taking a point charge as the source of the electromagnetic field
\begin{equation}
    J=Q\d(\vec{r})\pd_t,
\end{equation}
the solution of the Maxwell's equations \eqref{max} reads
\begin{equation}
    \phi=\frac{Q}{4\pi r},\label{coulomb}
\end{equation}
which is just the Coulomb potential.

For the gauge field configuration in \eqref{gauge}, the electromagnetic energy-momentum tensor takes the following form
\begin{equation}\label{Tel}
    T^\mu_{\ \nu}|_\text{el.} = \text{diag}\left\{-\frac{1}{2}{\f^\pr}^2,-\frac{1}{2}{\f^\pr}^2,\frac{1}{2}{\f^\pr}^2,\frac{1}{2}{\f^\pr}^2\right\}.
\end{equation}
Inserting the Coulomb potential into \eqref{Tel} and using the result in the Einstein equations \eqref{EM} together with the matter energy-momentum tensor [using (\ref{Tmatt}, \ref{ptoeps}) with $\epsilon = M \d(\vec{r})$] yield the metric function of the Reissner-Nordstr\"{o}m black hole as
\begin{equation}
    f(r) = 1 - \frac{2M}{r} + \frac{Q^2}{4\pi r^2}.
\end{equation}
Notice that the components of the electromagnetic energy-momentum tensor in \eqref{Tel} do not seem to satisfy the consistency condition
\begin{align}
    &T^\mu_{\ \nu}|_\text{el.} = \text{diag}\left\{-\e|_\text{el.}, 
    p_r|_\text{el.}, 
    p_\perp|_\text{el.}, 
    p_\perp|_\text{el.}\right\}\\
    &p_r|_\text{el.}=-\e|_\text{el.},~ p_\perp|_\text{el.}=-\e|_\text{el.}-\frac{r}{2}\e^\pr|_\text{el.}.\label{constonp}
\end{align}
However, reading off $\e|_\text{el.}=\frac{1}{2}{\phi^\pr}^2$ and $p_\perp|_\text{el.}=\frac{1}{2}{\phi^\pr}^2$ from \eqref{Tel}, one may solve for the scalar potential satisfying the consistency conditions from the relation $p_\perp|_\text{el.}=-\e|_\text{el.}-\frac{r}{2}\e^\pr|_\text{el.}$. Remarkably, this gives
\begin{equation}
    \phi=\frac{c}{r},\qquad c\text{: constant},\label{coulombunique}
\end{equation}
which means that Einstein equations in \eqref{EM} are consistent only for the Coulomb potential and the constant $c$ is fixed to be $\frac{Q}{4 \pi}$ from the Maxwell equations \eqref{max}.

For the non-commutative geometry inspired Reissner--Nordstr\"om black hole, the substitution rule \eqref{subs} should be applied to both the point mass and the point charge. We have already seen how the point mass should be treated in equations (\ref{Tmatt2}, \ref{Tmatt3}). The source of the electromagnetic part is smeared as
\begin{equation}
    J=\rho(r)\,\pd_t,\qquad \rho(r)=\frac{Q}{(4\pi\th)^{3/2}}e^{-r^2/4\th}.\label{smearedJ}
\end{equation}
Using this in Maxwell's equations \eqref{max} results in the following scalar potential and radial electric field
\begin{align}
    \phi&=\frac{Q}{4\pi^{3/2}r}\g(1/2, r^2/4\th),\label{phimod}\\
	 E&=-\phi^\pr=\frac{Q}{2\pi^{3/2}r^2}\g{(3/2, r^2/4\th)}.\label{Emod}
\end{align}
The form of the electromagnetic energy-momentum tensor remains the same as \eqref{Tel}, however it will produce a different contribution to the field equations since the smearing of the electric charge density modifies the scalar potential. From the $tt$-component of the Einstein equations \eqref{EM}, one finds the metric function as \cite{Ansoldi:2006vg}
\begin{equation}
    f(r) = 1 - \frac{4M}{\sqrt{\pi}r}\g(3/2,r^2/4\th)+\frac{Q^2}{4\pi^2r^2}\left(\g^2(1/2,r^2/4\th)-\frac{r}{\sqrt{2\th}}\g(1/2,r^2/2\th)\right).\label{NCRN}
\end{equation}
On the other hand, we have already proved that the only scalar potential satisfying the consistency condition for the components of the electromagnetic energy-momentum tensor in \eqref{Tel} is the Coulomb potential. Therefore, the scalar potential in \eqref{phimod} and the metric function in \eqref{NCRN} cannot satisfy the components of the Einstein equations where the transverse pressure is involved, which are the $\th\th$- and $\phi\phi$-components. An explicit calculation shows that this is indeed the case.

The failure can be exhibited in closed form and has a transparent physical origin. The energy-momentum tensor of Maxwell's theory is covariantly conserved only up to the Lorentz force acting on its source,
\begin{equation}
    \nab_\mu T^{\mu}_{\ \nu}|_\text{el.}=F_{\nu\lambda}J^{\lambda},\label{noncons}
\end{equation}
which vanishes for a point charge, whose current is zero everywhere except at the origin, but not for a smeared one. For the static configuration (\ref{gauge}, \ref{smearedJ}), the only nonvanishing component of the right-hand side of \eqref{noncons} is the radial one, with magnitude $\rho E$, which is the electrostatic self-force density of the smeared charge distribution. Since, for the metric ansatz \eqref{met}, the $tt$- and $rr$-components of the Einstein equations are automatically consistent with each other [by $G^t_{\ t}=G^r_{\ r}$ in \eqref{Einsteintensor} and $p_r=-\e$ in both sectors], the $\th\th$-component is the only independent check, and it is equivalent to the covariant conservation of the total source. The unbalanced force density therefore appears directly as the failure of the angular equations: evaluating the Einstein tensor of the metric \eqref{NCRN} against the tensors (\ref{Tmatt2}, \ref{Tmatt3}) and \eqref{Tel}, we find that the $tt$- and $rr$-components are satisfied, while
\begin{equation}
    G^{\th}_{\ \th}-8\pi\left(T^{\th}_{\ \th}|_\text{matt.}+T^{\th}_{\ \th}|_\text{el.}\right)=-4\pi r\,\rho E=-\frac{Q^2}{4\pi^2\th^{3/2}\,r}\,\g(3/2,r^2/4\th)\,e^{-r^2/4\th},\label{mismatch}
\end{equation}
and identically for the $\f\f$-component. Two features of \eqref{mismatch} explain how the inconsistency could remain unnoticed. First, it is invisible to everything that follows from the metric function alone: the $tt$-component correctly determines \eqref{NCRN}, so the horizon structure, geodesics and the thermodynamics derived in the literature are unaffected, and the defect can only surface in applications that make direct use of the energy-momentum tensor, such as the energy conditions studied below. Second, the mismatch is proportional to $e^{-r^2/4\th}$: it is exponentially suppressed away from the smearing region, so that any check of the field equations at or outside the horizon scale appears satisfied to high accuracy, and it vanishes identically in the commutative limit $r/\sqrt{\theta} \to \infty$, in which \eqref{Tel} with the Coulomb potential is consistent.

In order to satisfy these components, an improvement of the electromagnetic tensor is required such that the constraint on the transverse pressure $p_\perp|_\text{el.}$ in \eqref{constonp} is imposed by hand. In terms of the scalar potential, the improved electromagnetic energy-momentum tensor is given by
\begin{align}
	\cT^\mu_{\ \nu}|_\text{el.} &= \text{diag}\left\{-\e|_\text{el.},
	p_r|_\text{el.},
	p_\perp|_\text{el.},
	p_\perp|_\text{el.}\right\},\label{improvedT}\\
\e|_\text{el.} = \frac{1}{2}{\f^\pr}^2,& \quad
	p_r|_\text{el.}=-\e|_\text{el.}, \quad
	p_\perp|_\text{el.}= -\e|_\text{el.} - \frac{r}{2}\pd_r\e|_\text{el.}.\label{improvedT2}
\end{align}

The relation of the improved tensor to the standard construction \eqref{Telgen} can be stated precisely. The improved tensor is \emph{not} expressible in the form \eqref{Telgen}: as we have proven around \eqref{coulombunique}, on the ansatz \eqref{met} a tensor of that form satisfies the constraints \eqref{constonp} only for the Coulomb potential, which the smeared potential \eqref{phimod} is not. Instead, inserting \eqref{Emod} into (\ref{improvedT}, \ref{improvedT2}), the improved tensor decomposes as
\begin{equation}
    \cT^\mu_{\ \nu}|_\text{el.}=T^\mu_{\ \nu}|_\text{el.}-\frac{r}{2}\,\rho E\;\text{diag}\left\{0,0,1,1\right\},\label{decomp}
\end{equation}
i.e. the standard Maxwell tensor \eqref{Tel} supplemented by an interaction stress supported on the smeared current, whose covariant divergence cancels the Lorentz force \eqref{noncons} so that $\nab_\mu \cT^{\mu}_{\ \nu}|_\text{el.}=0$. Within the static diagonal ansatz this correction is unique: once the $tt$-component fixes the metric function, all components of any consistent total source are determined by the Einstein tensor, in particular
\begin{equation}
    8\pi\left(p_\perp|_\text{matt.}+p_\perp|_\text{el.}\right)=G^{\th}_{\ \th}.\label{uniqueness}
\end{equation}
Equivalently, one may keep the Maxwell tensor \eqref{Tel} untouched and assign the
correction in \eqref{decomp} to the matter sector,
$p_\perp|_\text{matt.}\to p_\perp|_\text{matt.}-\frac{r}{2}\rho E$: the smeared
charged matter then carries precisely the transverse stress required for its
equilibrium against its own electrostatic repulsion. The total energy-momentum
tensor, which is all that the Einstein equations are sensitive to, is identical in
the two descriptions.

Taking the trace of \eqref{decomp}, the Maxwell part drops out by conformal symmetry and one is left with the interaction term alone,
\begin{equation}
    \cT^\mu_{\ \mu}|_\text{el.}=-r\,\rho(r)\,E(r)=-\frac{Q^2}{16\pi^{3}\th^{3/2}\,r}\,\g(3/2, r^2/4\th)\,e^{-r^2/4\th}.\label{traceel}
\end{equation}
Adding the contribution of the matter sector (\ref{Tmatt2}, \ref{Tmatt3}), the trace of the total energy-momentum tensor reads
\begin{equation}
    \cT =\frac{1}{16\pi^{3/2}\th^{3/2}}\left (\frac{M(r^2-8\th)}{\th} - \frac{Q^2}{\pi^{3/2}r}\,\g(3/2, r^2/4\th) \right )e^{-r^2/4\th}\neq0.\label{trace}
\end{equation}
Therefore, the conformal symmetry of Maxwell's theory in four dimensions is broken
by the improvement that we propose. However, \eqref{traceel} shows that the breaking
is an interaction effect localized on the smeared source: it is exponentially
suppressed in the commutative limit $r/\sqrt{\th}\to\infty$, in which the improved
tensor reduces to the traceless Maxwell tensor \eqref{Tel} evaluated on the Coulomb
potential \eqref{coulomb}. All the components of the Einstein equations
$G^\mu_{\ \nu}= 8\pi\left({T}^\mu_{\ \nu}|_\text{matt.} + \cT^\mu_{\ \nu}|_\text{el.}\right)$
are now satisfied.
\begin{figure}
    \centering
    \includegraphics[width=\linewidth]{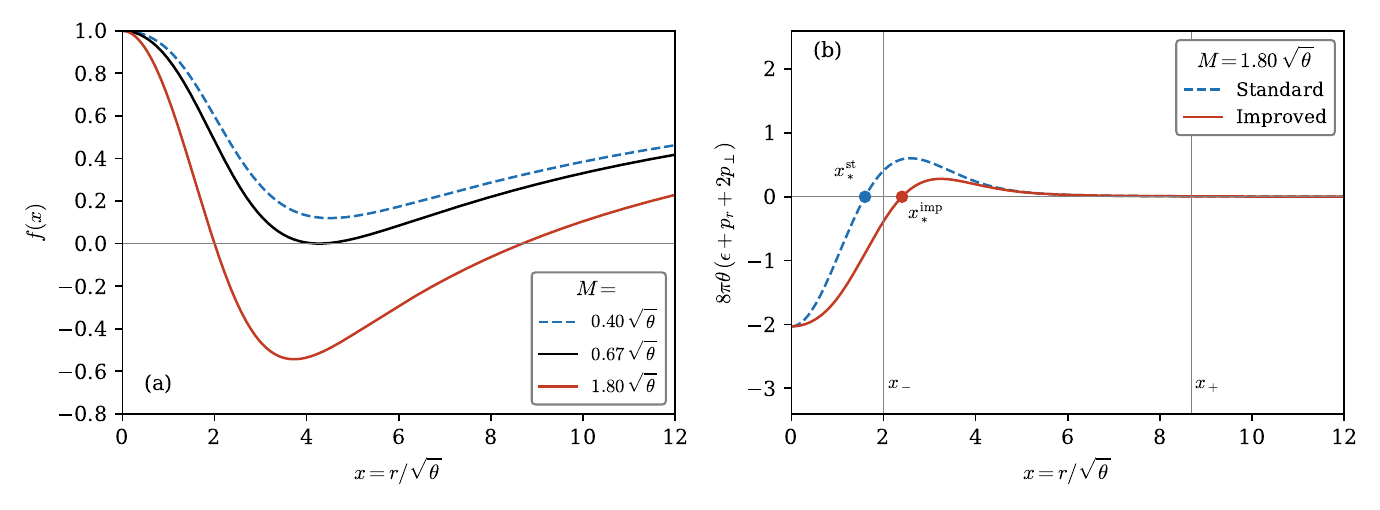}
    \caption{Metric functions and strong energy curves for the charge $Q = 15 \sqrt{\th}$. a) For different values of the mass $M$, metric functions are shown. In order to have two distinct event horizons, the mass should satisfy $M>M_0 = 0.67 \sqrt{\th}$. b) For $M = 1.80\sqrt{\th}$, the strong energy curves $8\pi\th\left(\e+p_r+2p_\perp\right)$ resulting from the standard and the improved energy-momentum tensors, possessing well-separated critical radii $x_*$. While the strong energy condition is violated only inside the Cauchy horizon for the standard case, $x_*^{\rm st}=1.60<x_-$, the improved energy-momentum tensor predicts a violation outside the Cauchy horizon, $x_*^{\rm imp}=2.40>x_-$.}
    \label{fig1}
\end{figure}

As mentioned earlier, many of the physical properties involving the thermodynamics of the solution are studied using the metric function and remain valid. However, for the type of energy-momentum tensors that we study, the strong energy condition reduces to the positivity of the transverse pressure as follows
\begin{equation}
    \e+p_r+2p_\perp \geq 0\quad\Rightarrow\quad p_\perp\geq0,
\end{equation}
since $p_r=-\e$. Therefore, the improved energy-momentum tensor that we propose predicts a different region where the energy condition is violated. Note that, by \eqref{uniqueness}, the critical radius obtained from the improved tensor is the unique one compatible with the Einstein equations: it is determined by the zero of $G^{\th}_{\ \th}$ and is therefore independent of how the total source is split between the matter and the electromagnetic sectors. In Figure 1, we present our results for $Q=15\sqrt{\th}$. In Figure 1(a), we show that the mass of the black hole should satisfy $M>M_0=0.67\sqrt{\th}$ in order to have two distinct event horizons $r_-$ and $r_+$. In Figure 1(b), the strong energy curves for the standard and the improved energy-momentum tensors are shown with the choice $M=1.80\sqrt{\th}$, for which the inner horizon is located at $r_-=2.00\sqrt{\th}$ and the outer is at $r_+ = 8.68 \sqrt{\th}$. For the former case, the energy condition is violated when $r<r_*=1.60\sqrt{\th}$. However, with the improved energy-momentum tensor, the violation occurs when $r<r_*=2.40\sqrt{\th}$. For the violation to be hidden inside a Cauchy horizon, we need to have $r_*<r_-$. Therefore, according to the standard energy-momentum tensor, which does not satisfy all the components of the field equations, the violation occurs inside the Cauchy horizon. However, there is violation outside the Cauchy horizon according to the improved energy-momentum tensor. For this value of the charge, the two treatments disagree qualitatively throughout the window $1.22\sqrt{\th}\lesssim M\lesssim 2.42\sqrt{\th}$; for smaller masses both place the violation inside the Cauchy horizon and for larger masses both place it outside, but the critical radii, and hence the extent of the violating region, differ for all values of the parameters. This clearly shows that certain parameter choices, although seemingly safe under the standard treatment, in fact place matter violating the strong energy condition in the region beyond the Cauchy horizon, which is relevant for the stability of the inner horizon and for strong cosmic censorship.

In this work, we revisited the non-commutative geometry inspired Reissner-Nordstr\"{o}m black hole solution obtained by smearing the point sources with a Gaussian distribution, which is motivated by the field theoretical results obtained by using coherent states. We showed that, while correctly reproducing the metric function from the $tt$-component of the Einstein equations, the standard procedure fails to satisfy all components due to the inconsistency of the resulting electromagnetic energy-momentum tensor. In particular, since the transverse pressure does not respect the constraint on the components of the energy-momentum tensor, $\th \th$- and $\f \f$-components of the Einstein equations are not satisfied. We exhibited the failure in closed form: it is exactly the Lorentz force density exerted by the electric field on the smeared charge, which the standard tensor leaves unbalanced, and it is exponentially suppressed away from the smearing region, which explains why it is invisible in all the metric-level properties of the solution studied so far.

As a remedy of this issue, we proposed an improved electromagnetic energy-momentum tensor that makes the Einstein equations consistent. The improvement is based on imposing the constraints on the energy-momentum tensor by hand, and it is the unique conserved completion of the standard Maxwell tensor: it is equal to the latter supplemented by an interaction stress supported on the smeared current. Although it breaks the conformal symmetry of Maxwell's theory in four dimensions, the breaking is localized on the smeared source and disappears in the commutative limit, and it guarantees the consistency of the field equations. This modification does not change the physical properties derived directly from the metric function such as geodesics or the thermodynamics of the solution. However, there might be significant changes when the energy-momentum tensor is involved in an application. As an explicit example, we demonstrated that the region, where the strong energy condition is violated, is quite sensitive to which energy-momentum tensor is used. For certain choice of the parameters of the black hole, the violation is confined within the Cauchy horizon according to the standard treatment. However, we observed a violation outside the horizon when our improved energy-momentum tensor is used.

This result suggests that the causal structure of non-commutative geometry inspired charged black holes may need to be reconsidered whenever the point electric charge is smeared. Since our improvement yields violations of the energy conditions outside the Cauchy horizon for some values of parameters, it might have important implications for the stability of the inner horizon and for strong cosmic censorship. As an alternative way to get rid of the inconsistency of Einstein equations without any manipulation by hand, one can also try to regularize the electric field by using a non-linear electromagnetic coupling. There are quite a number of examples in the literature including the celebrated Born-Infeld electrodynamics that give a regular electric field for a point charge source. Although contradicting the main philosophy of non-commutative geometry inspired gravitational solutions, which is using the substitution \eqref{subs} for any point sources, this seems to be an interesting possibility. We hope to report on this issue in future work.

\paragraph*{Acknowledgments} The authors acknowledge the use of Anthropic's AI assistant Claude, in particular for identifying the Lorentz force on the smeared charge as the origin of the inconsistency, and also for the verification of the expressions in the paper and the preparation of Figure \ref{fig1}. All results were independently verified by the authors. G. A. is supported by the Scientific and Technological Research Council of Turkey (T\"{U}B\.{I}TAK) under Grant Number 124F058.

\bibliographystyle{utphys}
\bibliography{ref}

\end{document}

%% file: pre.tex
\usepackage{latexsym,bm,graphicx,color,xcolor,nicefrac,titletoc,enumerate,amsmath,amssymb,xfrac,xcolor,physics,cite,setspace}

\usepackage{newtxtext,newtxmath,newtxtt}

\DeclareMathAlphabet{\mathcal}{OMS}{cmsy}{m}{n}

\usepackage[nottoc]{tocbibind} 
\usepackage{hyperref}
\hypersetup{linktocpage=true,colorlinks=true,linkcolor=blue,citecolor=blue,urlcolor=blue}
\usepackage[skip=3pt plus1pt, indent=20pt]{parskip}
\usepackage[bottom]{footmisc}
\usepackage{geometry}
\usepackage{titlesec}
\titleformat{\section}{\large\bfseries\sffamily}{\thesection}{0.5em}{}
\titleformat{\subsection}{\normalfont\bfseries\sffamily}{\thesubsection}{0.5em}{}
\titleformat{\subsubsection}{\normalsize\itshape\sffamily}{\thesubsubsection}{0.5em}{}
\titleformat*{\paragraph}{\normalsize\bfseries\sffamily}

\def\a{\alpha}
\def\b{\beta}
\def\g{\gamma}

\def\d{\delta}

\def\e{\epsilon}

\def\f{\phi}

\def\m{\mu}
\def\n{\nu}

\def\th{\theta}

\def\O{\Omega}

\def\pd{\partial}
\def\nab{\nabla}
\def\pr{\prime}

\newcommand{\cT}{\mathcal{T}}

\newcommand{\mail}[1]{\href{mailto:#1}{{\tt #1}}}